\documentclass[amssymb,prl,aps,twocolumn,amsmath,showpacs]{revtex4}

\usepackage{graphicx}

\begin{document}

\title{Density of states in a superconductor carrying a supercurrent}
\author{A.\ Anthore, H. Pothier, and D. Esteve}
\affiliation{Service de Physique de l'Etat Condens\'{e}, Direction des Sciences\\
de la Mati\`{e}re, CEA-Saclay, 91191 Gif-sur-Yvette, France}
\date{\today }

\begin{abstract}
We have measured the tunneling density of states (DOS) in a superconductor
carrying a supercurrent or exposed to an external magnetic field. The pair
correlations are weakened by the supercurrent, leading to a modification of
the DOS and to a reduction of the gap. As predicted by the theory of
superconductivity in diffusive metals, we find that this effect is similar
to that of an external magnetic field.
\end{abstract}

\maketitle

How is the superconducting order modified by a supercurrent? The superconducting order is based on pairing
electronic states which transform into one another by time reversal \cite{BCS}. The ground-state wavefunction
corresponds to a coherent superposition of doubly empty and doubly occupied time-reversed states, in an energy
range around the Fermi level given by the
BCS gap energy. When an external magnetic field $\vec{B}=\mathrm{curl}~\vec{A%
}$ is applied, time-reversed states are dephased differently, resulting in a
weakening of superconductivity. In presence of a supercurrent, the
superconducting order no longer corresponds to the pairing of time-reversed
states, which results in a kinetic energy cost, and again in a weakening of
superconductivity.\ In the early stages of the theory of superconductivity,
it was found that, in diffusive superconductors (in which the electron
mean-free-path is short compared to the BCS\ coherence length) and in
homogeneous situations, the modification of the superconducting order by a
magnetic field, by a current and by paramagnetic impurities could be
described by a single parameter, the depairing energy $\Gamma $ \cite{Parks}%
. Later on, the reformulation of the theory by Usadel \cite{Usadel,nes} in the diffusive limit extended this
equivalence to inhomogeneous situations, where the modulus of the order parameter may vary in space. In the Usadel equations, all physical quantities only involve the intrinsic combination $%
\overrightarrow{\nabla }\varphi -(2e/\hbar )\vec{A},$ where the gradient $%
\overrightarrow{\nabla }\varphi $ in the phase of the superconducting order
parameter is associated with the supercurrent, revealing the equivalence of
a supercurrent and of an applied magnetic field. The Usadel equations are
now at the basis of the understanding of mesoscopic superconductivity in
diffusive conductors \cite{Sophie,Courtois}. Experimentally, measurements of
the density of states (DOS) in a thin superconductor placed in an in-plane
magnetic field were well accounted for by the concept of depairing energy %
\cite{Levine}. In contrast, the effect of a supercurrent has been partly
addressed in a single experiment, focused on the reduction of the
superconducting gap close to the critical temperature \cite{Mitescu}. A
complication of the experiments with a supercurrent is that, if the sample
width exceeds the London penetration length $\lambda _{L}$, the current
distribution given by the non-local equations of electrodynamics \cite%
{Tinkham} is not homogeneous. In the experiment reported here, the
superconductor is wire-shaped, with thickness and width smaller than $%
\lambda _{L}$, so that the current flow is homogeneous and the magnetic field penetrates completely. Moreover,
the effect of the magnetic field induced by the supercurrent is then negligible. This simple geometry allows to
test the fundamental equivalence between the effect of a magnetic field and of a supercurrent in a diffusive
superconductor, and to compare precisely with the predictions of the Usadel equations.

\begin{figure}[bp]
\includegraphics[width=2.0in]{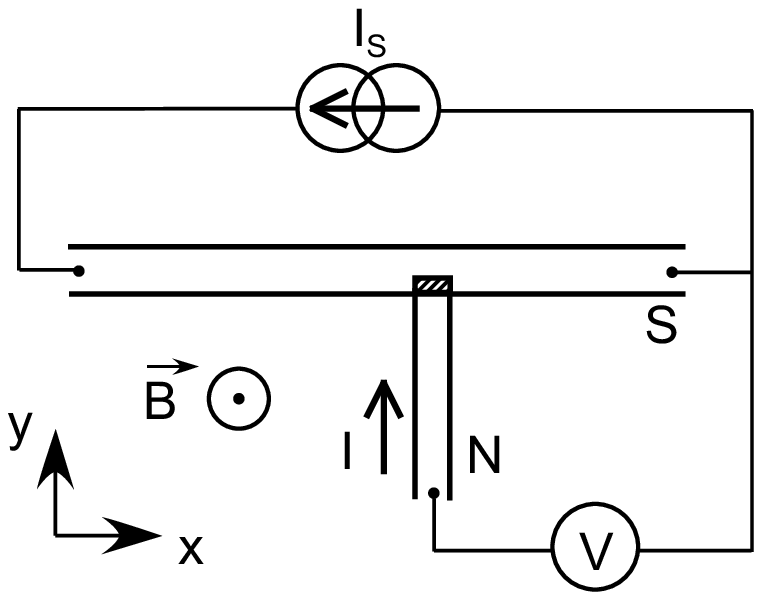}
\caption{Layout of the experiment: a $10~\protect\mu \mathrm{m}$-long, $120~%
\mathrm{nm}$-wide and $40~\mathrm{nm}$-thick superconducting (aluminum) wire can be current biased at $I_{S}$ or
exposed to a magnetic field $B$. A normal probe electrode forms a tunnel junction (dashed area) with the wire.
To a good approximation (see text), the differential conductance of the junction $dI/dV(V)$ is proportional to
the DOS in the superconductor.} \label{fig1}
\end{figure}

Our experiment was performed on a current-biased superconducting wire made of aluminum, placed in a
perpendicular magnetic field $B$ (see Fig. 1). The
density of states in the wire was infered from the differential conductance $%
dI/dV\left( V\right) $ of a tunnel junction formed between a small section
of the wire and\ a normal probe electrode made of copper. Disregarding
Coulomb blockade and temperature effects (see below), $dI/dV\left( V\right) $
is proportional to the DOS $n(eV)$. The sample was fabricated in an
electron-beam evaporator in a single pump-down, using three-angle
shadow-mask technique through a PMMA suspended mask patterned using e-beam
lithography \cite{Fred}. The substrate was thermally oxidized silicon. The $%
10~\mathrm{\mu m}$-long aluminum wire, with width $w=120~\mathrm{nm}$ and
thickness $t=40~\mathrm{nm,}$ was superficially oxidized in order to form a
tunnel barrier with the copper probe electrode overlapping it on an area $%
150\times 60$~$\mathrm{nm}^{2}$. The sample was mounted in a copper box
thermally anchored to the mixing chamber of a dilution refrigerator.
Measurements were performed at $25\mathrm{~mK}$. Electrical connections were
made through filtered coaxial lines. From the low-temperature,
high-magnetic-field wire resistance in the normal state, $R=77~\Omega ,$ the
conductivity $\sigma =27~\mathrm{\Omega }^{-1}\mathrm{.\mu m}^{-1}$ is
infered assuming that the electrical cross-section of the wire is $S=wt$.
The diffusion coefficient $D=49~\mathrm{cm}^{2}.\mathrm{s}^{-1}$ is then
deduced using Einstein's relation $\sigma =N(0)e^{2}D$, where $%
N(0)=2.15~10^{47}~\mathrm{J}^{-1}.\mathrm{m}^{-3}$ is the density of states
at the Fermi level of aluminum in its normal state and $e$ the electronic
charge.\textrm{\ }The superconducting gap $\Delta _{0}=205~\mathrm{%
%TCIMACRO{\U{b5}}%
%BeginExpansion
{\mu}%
%EndExpansion
eV}$ was deduced from the differential conductance-voltage characteristic $%
dI/dV\left( V\right) $ measured at $B=0,$ $I_{S}=0$ (dashed line in Fig.~2).
Using these parameters, we obtain the superconducting coherence length $\xi
_{0}=\sqrt{\hbar D/\Delta _{0}}\approx 125~\mathrm{nm}$ and the London
length $\lambda _{L}=\sqrt{\hbar /\left( \mu _{0}\pi \sigma \Delta
_{0}\right) }\approx 175~\mathrm{nm}$. Since $\lambda _{L}\gg w/2$, the
current density is homogeneous when the wire is current-biased, and a
magnetic field penetrates uniformly in the wire. The measured critical
current of the wire at $B=0$ was $I_{c}=106\mathrm{~%
%TCIMACRO{\U{b5}}%
%BeginExpansion
{\mu}%
%EndExpansion
A}$.

\begin{figure}[bp]
\includegraphics[width=3.0in]{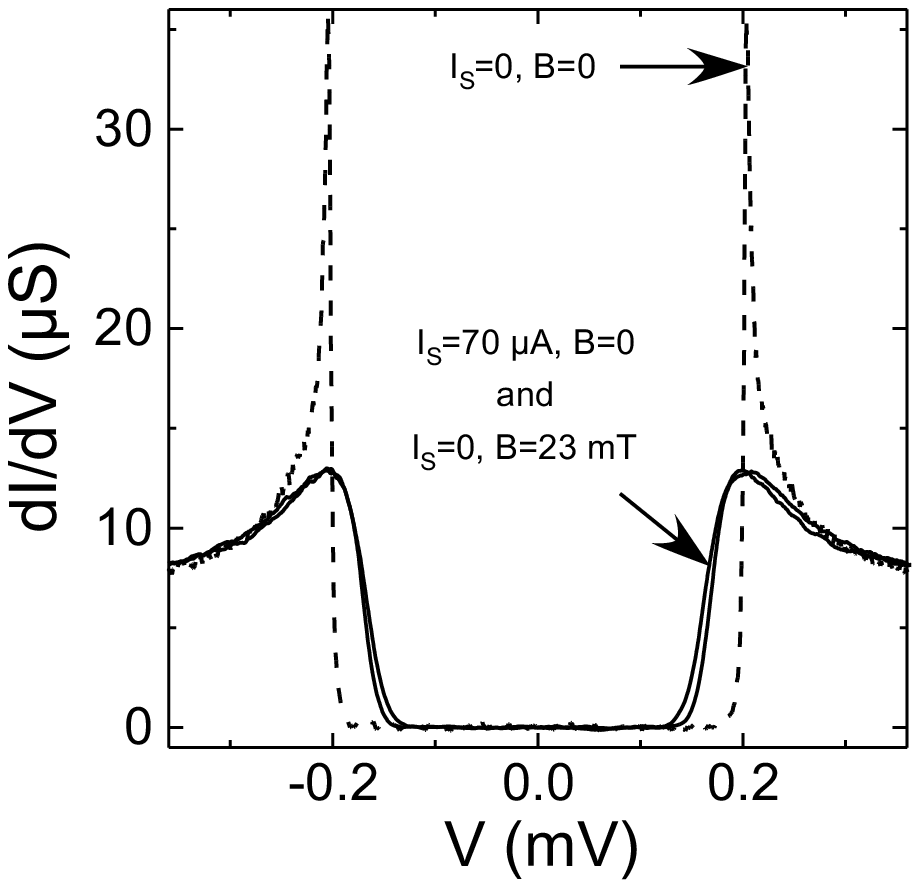}
\caption{Measured $dI/dV(V)$ for
different combinations of the bias current and magnetic field: dashed line: $%
I_{S}=0$ and $B=0$; solid lines: $I_{S}=70~\mathrm{%
%TCIMACRO{\U{b5}}%
%BeginExpansion
{\mu}%
%EndExpansion
A}$ and $B=0,$ and $I_{S}=0$ and $B=23~\mathrm{mT}.$} \label{fig2}
\end{figure}

In Fig. 2, two $dI/dV\left( V\right) $ curves are shown,
respectively measured at $I_{s}=70~\mathrm{%
%TCIMACRO{\U{b5}}%
%BeginExpansion
{\mu}%
%EndExpansion
A}$, zero field, and at zero current, $B=23~\mathrm{mT}$. The reduction of
the gap and the smearing of the peak near the gap energy are similar in the
two situations, bringing already evidence of the equivalent effect of $I_{S}$
and $B$. Note that the magnetic field created by the supercurrent has a
negligible effect: for $I_{s}=70~\mathrm{%
%TCIMACRO{\U{b5}}%
%BeginExpansion
{\mu}%
%EndExpansion
A}$ in the wire (see Fig.\ 2), $\frac{\mu _{0}I_{s}}{2\pi w}\sim 0.15~%
\mathrm{mT}$ whereas the resulting DOS\ is recovered at $I_{S}=0$ with $B=23~%
\mathrm{mT}$. A complete set of data is presented in Fig.~3, with $%
dI/dV\left( V\right) $ measured for $I_{S}=17,$ $51$ and $85~\mathrm{%
%TCIMACRO{\U{b5}}%
%BeginExpansion
{\mu}%
%EndExpansion
A}$ at $B=0$, and for $B=11.5$ to $69~\mathrm{mT}$ by steps of $11.5~\mathrm{%
mT,}$ at $I_{S}=0.$ Note that when the wire is current biased, the
superconducting state is metastable. In practice, for bias currents larger $%
85~\mathrm{%
%TCIMACRO{\U{b5}}%
%BeginExpansion
{\mu}%
%EndExpansion
A}$, the system switches to the resistive state during the recording of the $%
dI/dV(V)$ curve. The measured curve is then similar to that obtained in the
normal state. In order to account quantitatively for the data, we use the
Usadel theory \cite{Usadel,nes}. In this theory, correlations between
electrons of opposite spins and momenta are described by a complex function $%
\theta (\vec{r},E)$, the pairing angle, which depends both on space and
energy, and a local complex phase $\varphi (\vec{r},E)$. The local density
of states is given by $n(\vec{r},E)=N(0){\mathrm Re}[\cos (\theta (\vec{r},E))]$%
. The pairing angle and the complex phase obey the Usadel equations:

\begin{equation}
\frac{\hbar D}{2}\nabla ^{2}\theta +[iE-\frac{\hbar D}{2}(\overrightarrow{%
\nabla }\varphi -\frac{2e}{\hbar }\vec{A})^{2}\cos \theta ]\sin \theta
+\Delta \cos \theta =0  \label{usadel1}
\end{equation}

\begin{equation}
\overrightarrow{\nabla }[(\overrightarrow{\nabla }\varphi -\frac{2e}{\hbar }%
\vec{A})\sin ^{2}\theta ]=0.  \label{usadel2}
\end{equation}%
A term describing spin-flip scattering, which is found negligible in our
experiment, has been omitted here. The pairing potential $\Delta (\vec{r})$
is determined self-consistently by%
\begin{equation}
\Delta (\vec{r})=N(0)V_{eff}\int_{0}^{\hbar \omega _{D}}\mathrm{d}E\tanh (%
\frac{E}{2k_{B}T})\mathrm{Im}(\sin \theta )  \label{gap}
\end{equation}%
where $V_{eff}$ is the pairing interaction strength, $\omega _{D}$ the Debye
pulsation, $k_{B}$ the Boltzmann constant and $T$ the temperature of the
superconductor.

The supercurrent density$\ \vec{j}$ is given by:%
\begin{equation}
\vec{j}(\vec{r})=\frac{\sigma }{e}\int_{0}^{\infty }\mathrm{d}E\ \tanh (%
\frac{E}{2k_{B}T})\mathrm {Im}(\sin ^{2}\theta )(\overrightarrow{\nabla }%
\varphi -\frac{2e}{\hbar }\vec{A}).  \label{supercurrent}
\end{equation}%
In a situation like ours where the system consists entirely of a single
superconductor, $\overrightarrow{\nabla }\varphi $ does not depend on
energy, and $\ \vec{j}$ can be written as a product of the density of charge
in the superconducting state $\rho _{S}(\vec{r})=eN(0)U_{S}(\vec{r}),$ with $%
U_{S}(\vec{r})=\int_{0}^{\infty }\mathrm{d}E\tanh (\frac{E}{2k_{B}T})\mathrm {Im%
}(\sin ^{2}\theta ),$ and of a superfluid velocity $\mathcal{\vec{V}}_{S}=D(%
\overrightarrow{\nabla }\varphi -(2e/\hbar )\vec{A}).$

\begin{figure}[tbp]
\includegraphics[width=3.0in]{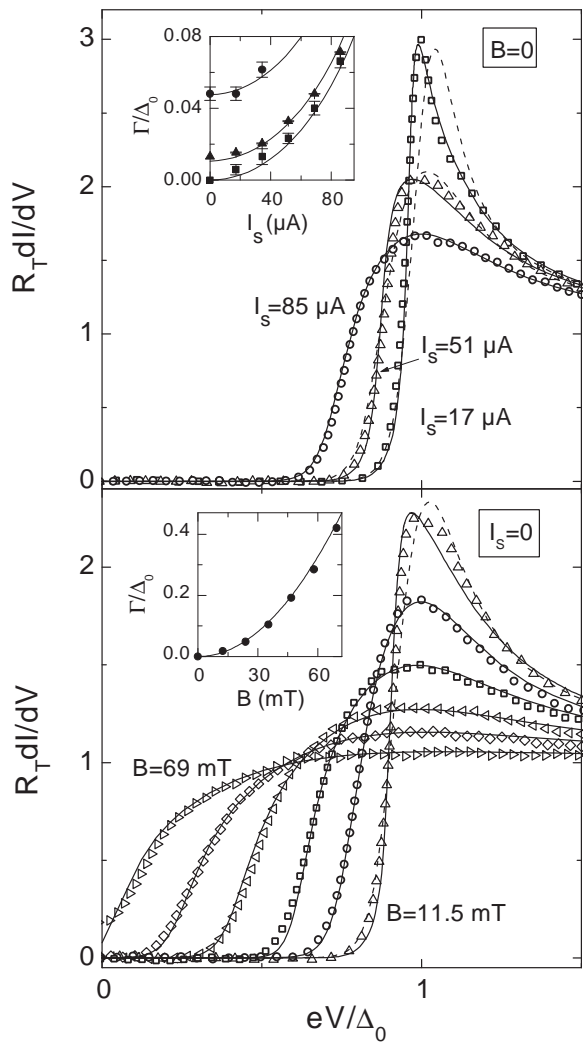}
\caption{Normalized differential conductance $%
dI/dV(V)$ of the probe tunnel junction: Top: at $B=0,$ as a function of the supercurrent $I_{S}$ (from right to
left: $I_{S}=17~\mathrm{\protect\mu A}$,
$51~\mathrm{%
%TCIMACRO{\U{b5}}%
%BeginExpansion
{\mu}%
%EndExpansion
A}$, and $85~\mathrm{\protect\mu A)}$. Bottom: at $I_{S}=0$, as a function of the magnetic field $B$ (from
$11.5~\mathrm{mT}$ to $69~\mathrm{mT}$ by steps of $11.5~\mathrm{mT)}$. Solid lines are best fits with
$dI/dV(V)$ calculated with an electronic temperature dependent on $V$ (see text); dashed lines are the best fits
with $dI/dV(V)$ calculated with a constant electronic temperature. Insets: depairing energy $\Gamma $(in units
of the gap $\Delta _{0}$ at $B=0$ and $I_{S}=0$) for different currents and magnetic fields, deduced from the
fits of $dI/dV(V).$ In the top inset, square symbols correspond to the data in the main panel ($B=0),$ whereas
triangles and disks were obtained from data taken at $B=10.2~\mathrm{mT}$ and $B=27~\mathrm{mT,}$ respectively.
Solid lines: Fits with theory, leading
to depairing current and magnetic field $I_{\Gamma }=240~\protect\mu \mathrm{%
A}$ and $B_{\Gamma }=105~\mathrm{mT.}$} \label{fig3}
\end{figure}

We have first checked numerically that the dependence of $\theta $ on the
directions transverse to the wire could be neglected because the width and
thickness are smaller than the superconducting coherence length $\xi _{0},$
which is the characteristic lengthscale for the variations of $\theta .$ As
a consequence, all the quantities can be replaced by their values averaged
on the transverse directions. In the London gauge, the effect of the
magnetic field is described by a vector potential parallel to the wire axis $%
x$, with an amplitude $A_{x}=By$, so that $\left\langle A_{x}\right\rangle
=0 $ and $\sqrt{\left\langle A_{x}^{2}\right\rangle }=Bw/(2\sqrt{3})$ \cite%
{penetre}. The constant phase gradient $\partial \varphi /\partial x$ is
given by the supercurrent $I_{S}=jS=U_{S}L/(eR)\left( \partial \varphi
/\partial x\right) $. Since $\partial ^{2}\varphi /\partial x^{2}=0$, Eq. (%
\ref{usadel2})\ reduces to $\partial (\sin ^{2}\theta )/\partial x=0$. No
spatial dependence remains in Eq. (\ref{usadel1}), and one recovers the
generic equation given in Ref. \cite{Parks}:

\begin{equation}
E+i\Gamma \cos \theta =i\Delta \frac{\cos \theta }{\sin \theta }
\label{usadelsimple}
\end{equation}%
where
\begin{equation}
\Gamma =\frac{\hbar D}{2}\left( \left( \frac{\partial \varphi }{\partial x}%
\right) ^{2}+\left( \frac{2e}{\hbar }\right) ^{2}\left\langle
A_{x}^{2}\right\rangle \right)
\end{equation}%
is the depairing energy, which contains the effect of both a phase gradient
and a magnetic field. Note that since $\Gamma /\Delta _{0}=$ $\frac{1}{2}%
(\xi _{0}\partial \varphi /\partial x)^{2}+\frac{1}{6}(\xi _{0}wB/(\hbar
/e))^{2}$ the relevant parameters are the phase difference between two
points of the wire distant by $\xi _{0}$ and the number of flux quanta in an
area $w\xi _{0}$. The depairing energy is related to the external parameters
$I_{s}$ and $B$ by the equation:%
\begin{equation}
\frac{\Gamma }{\Delta _{0}}=\left( \frac{\Delta _{0}}{U_{S}(\Gamma )}\frac{%
I_{s}}{I_{\Gamma }}\right) ^{2}+\left( \frac{B}{B_{\Gamma }}\right) ^{2},
\label{gammasurd}
\end{equation}%
where we have introduced the characteristic depairing current and magnetic
field $I_{\Gamma }=\sqrt{2}\Delta _{0}/(eR(\xi _{0})),$ with $R(\xi
_{0})=R\xi _{0}/L$ the resistance of the wire on a length $\xi _{0},$ and $%
B_{\Gamma }=\sqrt{6}(\hbar /e)/(w\xi _{0}).$ Since the transverse dimensions
of the wire are smaller than the London length $\lambda _{L},$ the depairing
energy due to the induced field is negligible (smaller by a factor $\sim
10^{-4}$ \cite{cylindre}) compared to the one due to the supercurrent. The
DOS for a given depairing energy $\Gamma $ is obtained from the
self-consistent solution of Eqs.~(\ref{gap}) and (\ref{usadelsimple}). For
practical purposes, we give the approximate expressions for the resulting $%
\Delta (\Gamma )/\Delta _{0}$ and $U_{s}(\Gamma )/\Delta _{0}$, valid, at $%
k_{B}T\ll \Delta ,$ for $\Gamma /\Delta _{0}\lesssim 0.3$:%
\begin{eqnarray}
\frac{\Delta (\Gamma )}{\Delta _{0}} &\simeq &1-0.75\frac{\Gamma }{\Delta
_{0}}-0.54\left( \frac{\Gamma }{\Delta _{0}}\right) ^{2}  \label{approx} \\
\frac{U_{s}(\Gamma )}{\Delta _{0}} &\simeq &\pi /2-1.8\frac{\Gamma }{\Delta
_{0}}-1.0\left( \frac{\Gamma }{\Delta _{0}}\right) ^{2}.  \nonumber
\end{eqnarray}

The differential conductance measured in the experiments is not exactly
proportional to the density of states $n(E)$ in the superconducting wire.
Two effects must be taken into account in order to calculate $dI/dV(V)$ from
$n(E):$ Coulomb blockade and the temperature of the probe electrode. Coulomb
blockade results from the finite impedance of the electromagnetic
environment of the tunnel junction \cite{Sophie}$.$ The characteristics of
the environment are found from the $dI/dV(V)$ characteristic of the circuit
in the normal state, reached at $B>0.1~\mathrm{T},$ which presents a 10\%
logarithmic dip at zero voltage. The environment can be modeled by a
capacitance $C=8~\mathrm{fF}$ in parallel with a resistance $R=250~\mathrm{%
\Omega } $. Coulomb blockade results in a convolution of the density of
states with a function $P(E),$ the probability for the electromagnetic
environment of the tunnel junction to absorb an energy $E$ \cite{cb}:
\begin{equation}
\frac{dI}{dV}(V)=\frac{1}{R_{t}}\int_{0}^{eV}\mathrm{d}E\ n(E)P(eV-E).
\label{cbeq}
\end{equation}%
Here, $P(E)=\alpha /E_{0}(E/E_{0})^{\alpha -1}$ for $E$ smaller than $%
E_{0}=e^{2}/\pi \alpha C,$ with $\alpha =2R/(h/e^{2}).$ The tunnel
resistance of the junction was $R_{t}=140~\mathrm{k\Omega }$. As a result of
this correction, the peak value of $n(E)$ is reduced by a few \% in $%
dI/dV(V).$ Finite temperature in the normal probe results in a further
convolution with the derivative of a Fermi function. In our experimental
setup, this temperature is slightly voltage-dependent, because the probe
electrode is thermally isolated from the larger contact pads by
superconducting connections. Heat transport only occurs by electron-phonon
coupling and by electron tunneling through the junction. Since both
mechanisms are very inefficient, even an input power $\mathcal{P}_{\mathrm{in%
}}$ in the fW range can induce a significant temperature increase. At bias
voltages large compared to the superconducting gap, heating by the tunneling
current has a sizeable effect. In contrast, at bias voltages $V$ slightly
below $\Delta /e$, only quasiparticules at energies larger than $\Delta -eV$
can tunnel, resulting in evaporative cooling \cite{John}. The effective
electron temperature $T$ is obtained by solving the heat equation:
\begin{equation}
\Sigma \Omega (T^{5}-T_{ph}^{5})-\mathcal{P}_{\mathrm{in}}+\int \mathrm{d}E\
\frac{E}{e^{2}R_{T}}\ n(E+eV)(1-f(E))=0.  \label{heat}
\end{equation}%
The first term describes heat transfer to the phonon bath, with $\Sigma
\simeq 2~\mathrm{nW}.\mathrm{\mu m}^{-3}.\mathrm{K}^{-5}$ for Cu \cite{Fred}%
, $\Omega \simeq 0.08~\mathrm{\mu m}^{3}$ the volume of the normal region of
the probe electrode, and $T_{ph}=25~\mathrm{mK}$ the phonon temperature. The
second term accounts for additional uncontrolled heat flow, that we
attribute to spurious electromagnetic noise. The third term accounts for
heat transfer throught the junction$,$ with $f(E)$ the Fermi function at
temperature $T$. From the fit of the data at $B=0$ and $I_{s}=0,$ we find $%
\mathcal{P}_{\mathrm{in}}=185~\mathrm{aW,}$ corresponding to $T=65~\mathrm{mK%
}$ at $eV\ll \Delta _{0}.$ The maximum cooling effect is reached at $%
eV/\Delta _{0}=0.99,$ where $T=30~\mathrm{mK;}$ heating dominates for $%
eV/\Delta _{0}>1.02,$ with $T=210~\mathrm{mK}$ at $eV/\Delta _{0}=1.5$.

In Fig. 3, we present with solid lines best fits of the data, taking into account both Coulomb blockade and
temperature corrections. The values of the fit parameter $\Gamma $ for each curve are given in the insets. For a
comparison, are also shown with dashed lines fits with a constant electron temperature ($T=60~\mathrm{mK}$). The
$V-$dependent temperature correction only matters for the sharpest curves. In turn, by fitting $\Gamma
(I_{s},B)/\Delta _{0}$ with Eq. (\ref{gammasurd}) and (\ref{approx})$,$ we find $I_{\Gamma }=240~\mu \mathrm{A}$
and $B_{\Gamma }=105~\mathrm{mT}.$ The theoretical values, assuming that the electrical dimensions of the wire
are
identical to the geometrical ones, are $I_{\Gamma }=310~\mu \mathrm{A}$ and $%
B_{\Gamma }=105~\mathrm{mT}$. Conversely, the experimental values of $%
I_{\Gamma }\propto \xi _{0}^{-1}$ and $B_{\Gamma }\propto (w\xi _{0})^{-1}$
can be used to extract effective values $\xi _{0\mathrm{eff}}=162~\mathrm{nm}
$ (instead of $125~\mathrm{nm}$) and $w_{\mathrm{eff}}=93~\mathrm{nm}$
(instead of $120~\mathrm{nm}$)\textrm{.} This corresponds in turn to an
increased value of the diffusive coefficient: $D=81~\mathrm{cm}^{2}\mathrm{s}%
^{-1}$ and, through the resistance, to an effective thickness $t_{\mathrm{eff%
}}=31~\mathrm{nm}$ (instead of $40~\mathrm{nm}$)$.$ Reduced effective
dimensions for electrical transport could be attributed partly to the
surface oxidation of the aluminum, which was exposed to air at atmospheric
pressure before measurement, and to surface roughness.

A by-product of the Usadel equations is a straightforward calculation of the
critical current. According to Eq. (\ref{supercurrent}), $I_{S}\propto
U_{s}(\Gamma )\partial \varphi /\partial x$. Since $U_{s}(\Gamma )$
decreases with $\Gamma $, $I_{s}$ presents a maximum as a function of $%
\partial \varphi /\partial x$, which is the thermodynamic critical current.
At $B=0$ and $k_{B}T\ll \Delta _{0},$ the maximum occurs at $\xi
_{0}\partial \varphi /\partial x\approx 0.69,$ and corresponds, in agreement
with \cite{Romijn}, to $I_{c}\approx 0.75S\Delta _{0}^{3/2}\sqrt{N(0)\sigma
/\hbar }\approx 0.53I_{\Gamma }=125~\mathrm{\mu A}$ (using the experimental
determination of $I_{\Gamma }$). The difference with the measured $I_{c}=106~%
\mathrm{\mu A}$ might be due to the uncontrolled environment of the wire and
to inhomogeneities in the wire cross-section.

In conclusion, we have measured by tunneling spectroscopy on a
superconducting wire the effect on the superconducting order of a
supercurrent $I_{S}$ and of an external magnetic field $B$. As predicted by
the theory of superconductivity in diffusive conductors, the overall effect
solely depends on a single parameter, the depairing energy $\Gamma $. For
our narrow wire, the Usadel equations lead to a simple expression for this
depairing energy as a function of $I_{S}$ and $B,$ which compares well with
the experimental determination of $\Gamma $.

We thank C. Mitescu for the communication of his PhD thesis and for
correspondance, and N.\ Birge for his comments on the manuscript. We
acknowledge the technical help of P. Orfila, and permanent input from M.
Devoret, P. Joyez, F. Pierre, C. Urbina, and D. Vion.


\begin{thebibliography}{99}
\bibitem{Levine} J.L.\ Levine, Phys. Rev. \textbf{155}, 373 (1967). In this
work, the films were thick enough to ensure that the deparing effect was
dominating the Zeeman effect.
\bibitem{penetre} When taking into account the screening of the magnetic
field inside the superconductor, we get a negligible correcting factor $%
\approx 1-1/60(w/\lambda _{L})^{2}\approx 0.99.$
\bibitem{cylindre} For a cylindrical wire with cross-area $S$, the ratio is $%
(S/\lambda _{L}^{2})^{2}/90.$
\end{thebibliography}
\end{document}